# Bacteria Identification by Phage Induced Impedance Fluctuation Analysis (BIPIF)

[1]Gabor SCHMERA and [2]Laszlo B. KISH

[1] Space and Naval Warfare Systems Center Pacific, San Diego, CA 92152-5001, USA
gabe.schmera@navy.mil

[2] Texas A&M University, Department of Electrical and Computer Engineering, College Station, TX 77843-3128, USA
laszlo@ece.tamu.edu

**Abstract:** We present a new method for detecting and identifying bacteria by measuring impedance fluctuations (impedance noise) caused by ion release by the bacteria during phage infestation. This new method significantly increases the measured signal strength and reduces the negative effects of drift, material aging, surface imperfections, 1/f potential fluctuations, thermal noise, and amplifier noise.

Comparing BIPIF with another well-known method, bacteria detection by SEnsing of Phage Triggered Ion Cascades (SEPTIC), we find that the BIPIF algorithm is easier to implement, more stable and significantly more sensitive (by several orders of magnitude). We project that by using the BIPIF method detection of a single bacterium will be possible.

## 1. Introduction: Fluctuation-enhanced sensing of chemicals and bacteria

Fluctuation-enhanced chemical [1-8] and biological [9-12] sensing (FES) utilizes the stochastic component of sensor signals that is caused by the statistical interaction between the agents and the sensor. A typical FES system utilizes specially designed sensors and advanced signal processing and pattern recognition algorithms [7-9].

In 2005 a method for detecting and identifying bacteria by SEnsing Phage Triggered Ion Cascades (SEPTIC) was proposed by Kish and coworkers [10-13]. The SEPTIC scheme is based on detecting and analyzing the electrical field (voltage) fluctuations caused by the stochastic emission of ions during phage infection. A two-electrode nano-well device is immersed in the carrier fluid containing a phage-infected sample and the microscopic voltage fluctuations are measured across the electrodes. In experiments the SEPTIC method could identify various strains of E. coli in less than 10 minutes [10]. The SEPTIC method also showed excellent specificity due to the specificity of the bacteriophages utilized.

However, the SEPTIC method has serious shortcomings that may prevent further development and commercialization. The method has not been shown to work for small bacteria samples; all experiments so far used large samples (typically 10 million bacteria) [10, 11, 12, 23]. Moreover, the observed power density spectral shapes showed significant variations when performed in different laboratories [10, 13].

In the present paper, we introduce a new method for identifying bacteria: Bacteria Identification by Phage Induced Impedance Fluctuation analysis (BIPIF). BIPIF is predicted to significantly outperform SEPTIC, thus it shows a potential for further development and commercialization.

The SEPTIC technique measures fluctuations in the DC electrical field, i.e., the separation of positive and negative ions is the underlying and assumed phenomenon. The BIPIF method, on the other hand, measures the changes in AC impedance; the separation of positive and negative ions is not needed, the method works even when the negative and positive ions are in balance.

The BIPIF method offers several orders of magnitude improvement in sensitivity and higher reproducibility at the expense of somewhat more sophisticated sensor circuitry and signal processing algorithms.

This paper is organized as follows. In Section 2 we briefly analyze the detection limits of the SEPTIC method using the available measurement data. In Section 3, we introduce the BIPIF method, and in Section 4, we provide theoretical comparison and analysis of the two methods (SEPTIC and BIPIF).

## 2. The limits of the SEPTIC technique

The SEPTIC method's sensitivity is limited by the presence of strong 1/f background noise, drift, aging of the electrode material, and dependence on surface effects and corrosion [11, 12]. The combined negative effect of these limiting factors is most likely responsible for the seemingly poor reproducibility of measured spectral density functions; Table 1 shows measured spectral exponents for phage infected and control sample published by Kish et al. [10] and Seo et al. [13]. This apparent lack of reproducibility requires further investigation.

**Table 1.** Spectral exponents.

|  | Spectral exponent in [10] | Spectral exponent in [13] |
|---|---|---|
| Phage infected sample | -2 | -0.9 |
| Control sample (not infected) | -1 | -0.1 |

In order to enhance sensitivity, it is essential to increase signal strength and to minimize the effect of noise sources, such as 1/f noise, thermal noise [10, 13], and amplifier noise. The 1/f noise, which is caused by the DC potential fluctuations in the vicinity of the electrodes, is the primary sensitivity limiting factor for the SPETIC algorithm [12].

In this paper, we propose that an ion-sensitive measurement based on AC impedance fluctuations can be of significant help in overcoming the afore mentioned limiting factors. By using an AC driving current, signal strength will significantly increase, furthermore, the AC probing frequency can be much higher (such as 10 kHz) than the frequency range (1-10Hz) utilized by the SEPTIC method. The higher probing frequency will reduce the relative strength of the 1/f background noise by several orders of magnitude. In addition, both the thermal and amplifier noise interferences can also be drastically reduced utilizing sufficiently large AC current and two separate measuring frequencies with cross-correlation measurements. We project that the BIPIF method will significantly enhance reproducibility due to reduced noise interference.

## 3. The proposed new method: BIPIF and its advantages

Figure 1 shows the outline of the new BIPIF sensing system utilizing AC impedance fluctuation measurements. In this setup the interference from 1/f noise in the electrical Coulomb field at the electrode surfaces can be avoided by using relatively high probing frequency (such as 10 KHz) [16]. Furthermore, using two separate frequencies, a sufficiently large AC drive current, and utilizing cross-correlation measurements, the negative effects of the thermal noise and amplifier noise can also be reduced. By fine-tuning these system parameters, detecting a single infected bacterium becomes a possibility.

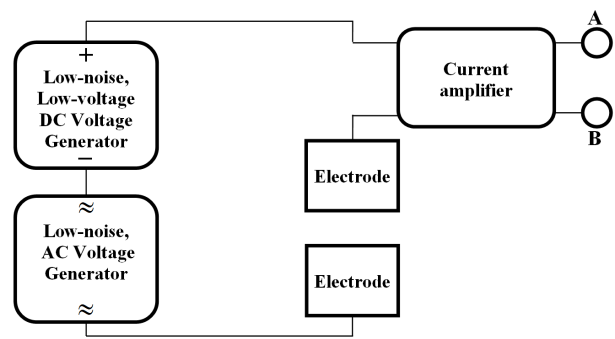

**Fig. 1.** A simple realization of the BIPIF method: a 2-electrode sensing system with current amplifier. The role of the DC voltage is to collect the bacteria to one of the end-electrodes.

To measure the impedance (conductance) between the two electrodes, an AC current is applied with high enough frequency such that the interference from the 1/f potential noise becomes negligible [16]. The AC current can be monitored either directly (Figure 1) or a three electrode bridge arrangement can be utilized (Figure 2) to measure the impedance difference across the electrodes.

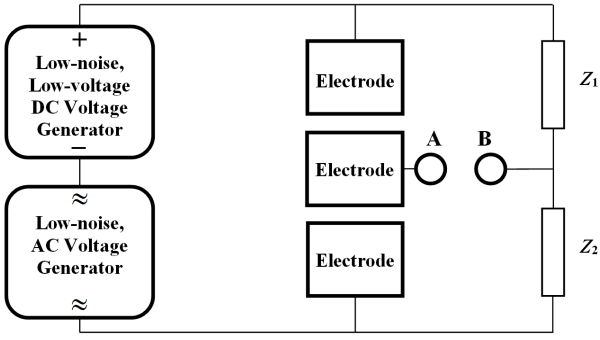

Fig. 2. A three-electrode bridge realization of the BIPIF method. The resistances Z1 and Z2 should be relatively small to keep thermal noise reasonably low.

Other similar arrangements with more than three electrodes are possible. Typically, the output voltage (across connection points A and B) connected to a preamplifier (not shown) and then to the differential input of a lock-in amplifier (Figure 3) driven by the same AC voltage generator that is connected to the electrodes.

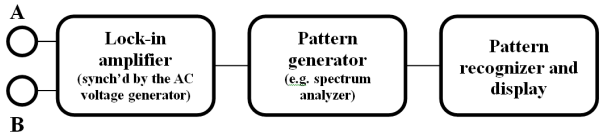

Fig. 3. Signal processing chain for BIPIF with one AC generator.

Utilizing the AC voltage generator and measuring the conductance fluctuations across the electrodes result in a significantly higher sensitivity compared to the SEPTIC method where passive spontaneous AC field fluctuations are measured. By properly setting the time-constant of the lock-in amplifier, its output will provide a slowly fluctuating AC signal component that is proportional to the low-frequency conductance fluctuations of the sample. In the processing chain, shown in Figure 3, the lock-in amplifier is followed by a pattern generator (for example, a spectrum analyzer) and then a pattern recognizer and a display.

In order to further improve the performance of the system, it is desirable to reduce the interference caused by the thermal noise and amplifier noise. This can be achieved by using two AC generators with different frequencies (Figure 4). In this arrangement two lock-in amplifiers are needed and the pattern generation is based on cross-correlation effects (cross-spectrum generation for instance).

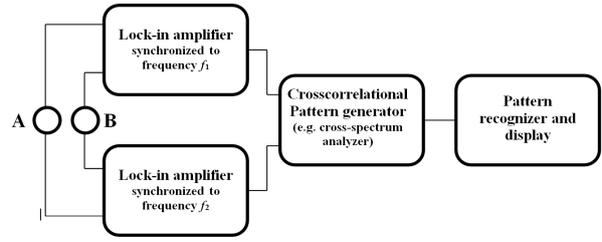

Fig. 4. Signal processing chain for BIPIF using two AC generators with different frequencies; two lock-in amplifiers used and the pattern generation is based on cross-correlation effects (cross-spectrum generation for instance).

## 4. Theoretical comparison between the BIPIF and SEPTIC methods

In order to quantitatively estimate the improvement in sensitivity by the proposed new method, we first analyze and compare the signal strengths produced by both methods, and then we examine how the presence of 1/f noise (and thermal noise) limits the sensitivity of both systems.

The SEPTIC method is based on a concentration cell (two electrodes of identical metals with fluctuating electrolyte concentration). The voltage $U_{cc}$ generated by a concentration cell is described by the Nernst equation [17]:

$$U_{cc} = \frac{kT}{Zq} \ln \frac{n_2}{n_1}, \qquad (1)$$

where k is the Boltzmann constant, T is the absolute temperature, Z is the valence number of the ions, q is the charge of an electron, and $n_1$ and $n_2$ are the ion concentrations in the vicinity of the electrodes. At room temperature (300 K), Eq. 1 reduces to:

$$U_{cc} = \frac{0.26}{Z} \ln \frac{n_2}{n_1} \; [\text{Volt}]. \qquad (2)$$

Let $n_2 = n_1 + \Delta n$ be the change in concentration at electrode 2 caused by a phage infestation. Assuming small relative concentration change, $|\Delta n| \ll n_1$, the observed voltage fluctuation (around the mean value) during SEPTIC measurement is:

$$\Delta U_{sep} = \frac{kT}{Zq} \ln \left( \frac{n_1 + \Delta n}{n_1} \right) = \frac{kT}{Zq} \ln \left( 1 + \frac{\Delta n}{n_1} \right) \\ \approx \frac{kT}{Zq} \frac{\Delta n}{n_1} = \frac{0.026}{Z} \frac{\Delta n}{n_1}. \qquad (3)$$

Let us now estimate the voltage fluctuations when using the BIPIF method. Here too, the ion concentrations in the vicinity of the electrodes will determine the conductance and its fluctuations even under anisotropic conditions [18]. For sake of simplicity, we assume that a single AC current generator is used; then the observed voltage amplitude fluctuations (around the mean value) that are due to conductance fluctuations during BIPIF measurement is simply:

$$\Delta U_{bip} = U_0 \frac{\Delta n}{n_1}, \quad (4)$$

assuming that the electrodes are approximately the same size. It is evident from equations (3) and (4) that characteristics of the signals measured by the two methods are very similar. However, the BIPIF method produces significantly higher signal levels (and drastically reduced noise levels as we'll see later in this section).

We measure the improvement or gain (G) in signal strength (power) by the squared ratio of the measured voltage fluctuations for the BIPIF and SEPTIC methods:

$$G = \left(\frac{\Delta U_{bip}}{\Delta U_{sep}}\right)^2 = \left(\frac{U_0 Z}{0.026}\right)^2. \quad (5)$$

As a concrete example, let's consider magnesium ions (Z=2) and 1 $V$ effective AC voltage ($U_0 = 1.41\ V$) drop between the electrodes (this value is proven to give Ohmic response with electrolytes [16]), then we obtain the gain:

$$G = \left(\frac{\Delta U_{bip}}{\Delta U_{sep}}\right)^2 = \left(\frac{1.41*2}{0.026}\right)^2 > 11700. \quad (6)$$

Thus the BIPIF signal power is increased by four orders of magnitude.

Let's now investigate the detection limits for both methods by examining interference from 1/f and thermal noise sources (while assuming temporarily that the signal level is the same for both methods). We assume linear sensor response w.r.t. the power density spectrum vs. the number of bacteria. This assumption is justified as long as the individual bacteria act as independent sources of fluctuations. Our argument here is a modified and improved version of the one presented in [12].

Figure 5 shows the measured power spectrum response of the SEPTIC system detecting E. coli bacteria using two different types of bacteriophages, T5 and Ur-λ. In this case, the 1/f noise is the limiting factor. Applying our linear response assumption, we find that the estimated sensitivity limit is ~30,000 bacteria using T5 phages, and ~1 million bacteria using Ur-λ phages.

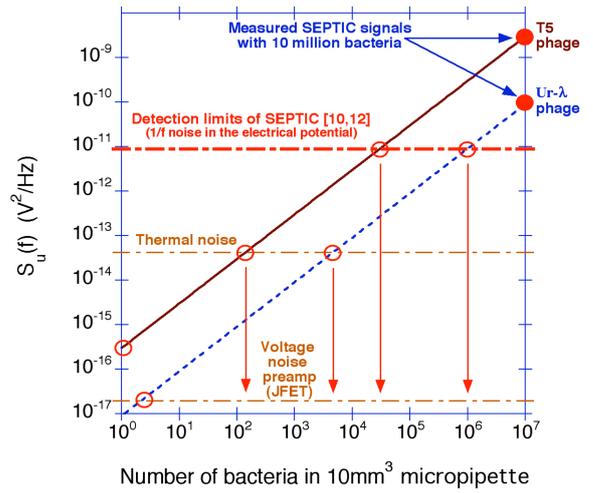

**Fig. 5.** Sensitivity limits for the SEPTIC and BIPIF methods. For the SEPTIC method the 1/f noise level determines the actual sensitivity limits of ~30,000 and ~1,000,000 bacteria assuming linear response. For the BIPIF method, sensitivity is limited by the Thermal noise level. The BIPIF method potentially improves sensitivity by three orders of magnitude due to eliminating 1/f noise as a limiting factor. Further improvement in the signal to noise ratio is possible by suppressing white noise via using two AC generators at different frequencies and cross-correlating pattern generation. This figure is based on Figure 2 of [12].

The BIPIF method suppresses the interference from the 1/f potential fluctuations and thus sensitivity is limited by thermal noise. Applying linear response assumption again we conclude that the BIPIF system's sensitivity is approximately three orders of magnitudes better due to 1/f noise suppression (Fig. 5); combining this result with the signal strength gain (Eq. 6) we see that the BIPIF will improve sensitivity by up to 7 orders of magnitude, thus detecting a single bacterium may become a possibility.

Further improvement in sensitivity is possible by reducing the interference from white noise sources such as thermal noise and amplifier noise. This can be achieved by using two AC generators with different frequencies (Figure 4). In this arrangement two lock-in amplifiers are needed and the pattern generation is based on cross-correlation effects (cross-spectrum generation for instance).

## 5. Conclusions

We introduced BIPIF, a new method of phage-based bacteria sensing and identification. A BIPIF-based sensing system is actively driven by one or more AC voltage generators and it measures the AC impedance fluctuations caused by ion release during phage infestation. We compared BIPIF's principles and effectiveness to SEPTIC's and showed that the new method performs significantly better by increasing signal strength, and by reducing or eliminating the negative effects of drift, material aging, surface imperfections, 1/f potential fluctuations, and thermal and amplifier noise. The BIPIF algorithm dramatically improves sensitivity; even detecting a single bacterium becomes a possibility.

## Acknowledgements

This research was carried out under the Navy Cooperative Research and Development Agreement (NCRADA- SSC Pacific – 08 – 116) between SPAWARSYSCEN-Pacific and Texas A&M University.

This paper includes material that is subject to a government-owned patent application.